\documentclass{Interspeech2024}
\usepackage{multirow}
\usepackage{tabularx}
\usepackage{pgfplots}
\pgfplotsset{height=4cm, width=8cm, compat=1.18}

\newcommand{\q}{\mathbf{q}}

\interspeechcameraready

\title{Differentiable Time-Varying Linear Prediction \\ in the Context of End-to-End Analysis-by-Synthesis}

\name{Chin-Yun}{Yu}
\name{György}{Fazekas}

\address{Queen Mary University of London, UK}
\email{chin-yun.yu@qmul.ac.uk, george.fazekas@qmul.ac.uk}

\keywords{speech synthesis, linear prediction, differentiable DSP, source-filter model, harmonic-plus-noise model}

\begin{document}

\maketitle

\begin{abstract}
Training the linear prediction (LP) operator end-to-end for audio synthesis in modern deep learning frameworks is slow due to its recursive formulation. In addition, frame-wise approximation as an acceleration method cannot generalise well to test time conditions where the LP is computed sample-wise. Efficient differentiable sample-wise LP for end-to-end training is the key to removing this barrier. We generalise the efficient time-invariant LP implementation from the GOLF vocoder to time-varying cases. Combining this with the classic source-filter model, we show that the improved GOLF learns LP coefficients and reconstructs the voice better than its frame-wise counterparts. Moreover, in our listening test, synthesised outputs from GOLF scored higher in quality ratings than the state-of-the-art differentiable WORLD vocoder.
\end{abstract}

\section{Introduction}

Efficiency and interpretability are important aspects of neural voice synthesis.
Instead of building high-quality but computationally expensive black-box vocoders~\cite{shen_natural_2017, kong2020hifi, prenger2019waveglow, kongDiffWaveVersatileDiffusion2021}, efforts have been made to increase controllability and reduce computational complexity while retaining the same voice quality. 
One promising approach is utilising synthesis components in the classic source-filter framework to split the workload of neural networks and train them jointly~\cite{valin2019lpcnet, wang2018neural}.
The filter in this framework represents the response of the vocal tract, which is often modelled using linear prediction (LP) based on the varying diameters tube model~\cite{markel_linear_1976}.
The LP coefficients (LPCs) are very compact features and require low transmission bandwidth.

Although LP has a long history in voice modelling~\cite{markel_linear_1976}, its recursive computation makes end-to-end training extremely slower than non-recursive filters due to the overhead for building deep computational graph~\cite{yu_singing_2023}.
LPCNet series~\cite{valin2019lpcnet, subramani_end--end_2022} address this problem by modelling the inverse filtered speech, thus utilising parallel computation.
Several works parallelise LP by frame-wise processing and overlap-add~\cite{mv_sfnet_2020, schulze2023unsupervised}, while Yu et al.~\cite{yu_singing_2023} accelerate it further with custom kernels for gradient backpropagation.
Other works seek to approximate the vocal tract filter using FIRs~\cite{juvela2019gelp, liu_neural_2020, wu1_ddsp-based_2022, nercessian2023differentiable}.
Besides LPCs, different filter representations have been explored as well, such as linear/mel-scale spectral envelope~\cite{wu1_ddsp-based_2022, nercessian2023differentiable} or cepstral coefficients~\cite{liu_neural_2020, yoshimura2023embedding}.

This paper proposes a new differentiable vocoder based on the GOLF vocoder~\cite{yu_singing_2023}.
We extend their custom backpropagation method to work with time-varying LP, removing mismatches between training and evaluation conditions with the cost of slightly slower training speed than frame-wise approximation.
We conducted an end-to-end analysis-by-synthesis experiment and compared the performance of several differentiable components with two classic synthesiser formulations.
We also compare the spectral envelopes of different methods' learnt LPCs.

\section{Background}
\subsection{Harmonic-plus-noise GOLF}
\label{ssec:hpn_golf}
GlOttal flow LPC Filter (GOLF) is a singing vocoder proposed by Yu et al.~\cite{yu_singing_2023}.
Given mel-spectrograms as conditions, a neural network encoder converts them into synthesis parameters, which are then passed to the decoder, also known as a synthesiser.
The synthesiser has the following form:
\begin{equation}
\label{eq:glottal_hpn}
    S(z) \equiv G(z)H(z) + N(z)C(z),
\end{equation}
where $S(z)$ is the voice signal which consists of $G(z)$ the glottal pulses filtered by a LP filter $H(z)$ and a white noise $N(z) \sim \mathcal{N}(0, 1)$ controlled by another LP filter $C(z)$.
This resembles the classic harmonic-plus-noise (HpN) model~\cite{serra_spectral_1990}, which has been used in several related works~\cite{liu_neural_2020, wu1_ddsp-based_2022, fabbro_speech_2020} with different harmonics and noise components.
GOLF generate $G(z)$ using wavetable synthesis, where the wavetables are sampled from the transformed LF model~\cite{fant_lf-model_1995} with different $R_d$. 

\subsection{Differentiable time-invariant LP}
\label{ssec:diff_ti_lp}
Although the response of the vocal tract varies from time to time in speech, we can assume it is invariant in a short time.
Based on this assumption, the computation of the LP filter is given by
\begin{equation}
\label{eq:lti_lpc} 
\begin{split}
    s(t) &= \text{LP}_{\bf a}\left(e(t)\right) \\
         &= e(t) - \sum_{i = 1}^M a_i s(t-i),
\end{split}
\end{equation}
where $s(t)$ is the time-domain speech signal, $e(t)$ the excitation signal, $\mathbf{a} = [a_1,\dots,a_M]$ the LP coefficients with a chosen order $M$.
$e(t)$ and $s(t)$ are zero for $t < 0$.
Yu et al. and Forgione et al.~\cite{yu_singing_2023, forgione_dynonet_2021} show that given a loss function $\mathcal{L}$ and the gradients $\frac{\partial \mathcal{L}}{\partial s(t)}$, the gradients $\frac{\partial \mathcal{L}}{\partial e(t)}$ and $\frac{\partial \mathcal{L}}{\partial a_i}$ can be computed as
\begin{gather}
\label{eq:lti_grad}
    \frac{\partial \mathcal{L}}{\partial a_i} = \sum_{t=0}^T \frac{\partial \mathcal{L}}{\partial s(t)}\text{LP}_{\bf a}\left(-s(t-i)\right) \\
    \frac{\partial \mathcal{L}}{\partial e(T - t)} = \text{LP}_{\bf a}\left(\frac{\partial \mathcal{L}}{\partial s(T - t)}\right),
\end{gather}
where $T+1$ is the number of audio samples and $T-t$ means reverse indexing. 
As the computation only consists of the same filter and matrix multiplication, one can register non-differentiable but efficient LP implementation into the computational graph to speed up gradient backpropagation in deep learning frameworks.

\section{Methodology}

\subsection{Source-filter GOLF}

Our first improvement to GOLF is using the following form:
\begin{equation}
\label{eq:glottal_sf}
    S(z) \equiv \left(G(z)+ N(z)C(z)\right)H(z),
\end{equation}
which closely resembles the source-filter (SF) model in~\cite{lu_glottal_nodate}, giving more explainability to the model, and has been used in several works~\cite{mv_sfnet_2020, nercessian2023differentiable, yoshimura2023embedding}.
Also, we empirically found that this form provides a more stable training curve and lower training loss than~\eqref{eq:glottal_hpn}.
This simple form is sufficient to model both voiced and unvoiced sounds by controlling the noise filter magnitude $|N(z)|$.
We use the FIR filter from~\cite{wu1_ddsp-based_2022} as $C(z)$, while $H(z)$ is our proposed LP filter with an input gain.

\subsection{Differentiable time-varying LP}
\label{ssec:diff_tv_lp}
Given  $\Tilde{\bf a}(t) = [\Tilde{a}_1(t),\dots,\Tilde{a}_M(t)]$ as time-varying filter coefficients, the sample-wise LP filter is:
\begin{equation}
\label{eq:ltv_lpc}
\begin{split}
    s(t) &= \text{LP}_{\Tilde{\bf a}(t)}\left(e(t)\right) \\
         &= e(t) - \sum_{i = 1}^M \Tilde{a}_i(t) s(t-i).
\end{split}
\end{equation}
Yu et al.~\cite{yu_singing_2023} divide the signal into short and overlapping frames and perform time-invariant LP independently on each frame to approximate~\eqref{eq:ltv_lpc}.
However, this method does not guarantee that the learnt coefficients generalise well after removing the approximation.
The frame size and overlap ratio also affect synthesis quality and must be chosen carefully.
We solved these issues by extending differentiable time-invariant LP (Sec.~\ref{ssec:diff_ti_lp}) to time-varying cases.

\subsubsection{The gradients to $e(t)$}

Eq.~\eqref{eq:ltv_lpc} equals filtering $e(t)$ with time-varying infinite impulse responses (IIRs) $\mathbf{b}(t) = [b_1(t),b_2(t),\dots]$ as
\begin{gather}
    s(t) = e(t) + \sum_{d=1}^t b_d(t) e(t-d) \label{eq:ltv_lpc_iir}\\
    b_d(t) = \sum_{\q \in \mathcal{G}_d} (-1)^{|\q|} \prod_{j=1}^{|\q|} \Tilde{a}_{q_j}\left(t - \sum_{k=1}^{j} Q(\q)_k\right) \label{eq:a2b}\\
    \mathcal{G}_d = \bigcup_{i=1}^{\min(d, M)} \left\{[i; \q]: \q \in \mathcal{G}_{d-i} \right\}, \label{eq:G_d}
\end{gather}
where $Q(\q) = [0; \q]$.
As the system is causal, the gradients to $e(t)$ depend on the future frames $s(>t)$ and can be expressed as
\begin{equation}
\label{eq:grad_e_iir}
\begin{split}
    \frac{\partial \mathcal{L}}{\partial e(t)} &= \frac{\partial \mathcal{L}}{\partial s(t)} \frac{\partial s(t)}{\partial e(t)} 
    + \sum_{d=1}^{T-t} \frac{\partial \mathcal{L}}{\partial s(t+d)} \frac{\partial s(t+d)}{\partial e(t)}\\
    &= \frac{\partial \mathcal{L}}{\partial s(t)} + \sum_{d=1}^{T-t} b_d(t+d) \frac{\partial \mathcal{L}}{\partial s(t+d)},
\end{split}  
\end{equation}
which means filtering the gradients backwards in time with shifted $b_d(t)$.
The issue is how to represent the filter in recursive form so we can reuse $\text{LP}_{\Tilde{\bf a}(t)}$ to reduce computational cost.
Let us plug $b_d(t+d)$ into~\eqref{eq:a2b} and get
\begin{equation}
\label{eq:b2a}
\begin{split}
    b_d(t+d) = \sum_{\q \in \mathcal{G}_d} (-1)^{|\q|} \prod_{j=1}^{|\q|} \Tilde{a}_{q_j}\left(t+d- \sum_{k=1}^{j} Q(\q)_k\right) \\
             = \sum_{\q \in \mathcal{G}_d} (-1)^{|\q|} \prod_{j=1}^{|\q|} \Tilde{a}_{q_j}\left(t+\sum_{k=j}^{|\q|} q_k\right) \\
             = \sum_{\Tilde{\q} \in \mathcal{G}_d} (-1)^{|\Tilde{\q}|} \prod_{j=1}^{|\Tilde{\q}|} \hat{a}_{\Tilde{q}_j}\left(t+\sum_{k=1}^{j} Q(\Tilde{\q})_k\right),
\end{split}
\end{equation}
where $\Tilde{\q} = [q_{|\q|},\ldots,q_1]$, $\hat{a}_i(t) = \Tilde{a}_i(t+i)$.
Comparing~\eqref{eq:b2a} to~\eqref{eq:a2b} after replacing $\Tilde{\q}$ with $\q$ (valid as both belong to $\mathcal{G}_d$), we observe that the IIRs $b_d(t+d)$ can be computed by applying $\hat{\bf a}(t) = [\hat{a}_1(t),\dots,\hat{a}_M(t)]$ LP backwards ($t = T \rightarrow t = 0$ due to changing the minus sign to plus sign).
Utilise this finding and~\eqref{eq:grad_e_iir}, we can express the gradients as
\begin{equation}
\label{eq:grad_e_lp}
    \frac{\partial \mathcal{L}}{\partial e(T - t)} = \text{LP}_{\hat{\bf a}(T-t)}\left(\frac{\partial \mathcal{L}}{\partial s(T - t)}\right).
\end{equation}

\subsubsection{The gradients to $\Tilde{\bf a}(t)$}

Let $z_i(t) = -\Tilde{a}_i(t)s(t-i)$ and express $s(t)$ as $e(t) + z_1(t) + \cdots + z_M(t)$.
The chain rule tells us that $\frac{\partial \mathcal{L}}{\partial e(t)} = \frac{\partial \mathcal{L}}{\partial z_i(t)}$ because $\frac{\partial s(t)}{\partial e(t)} = \frac{\partial s(t)}{\partial z_i(t)} = 1$.
Moreover, we already have $\frac{\partial \mathcal{L}}{\partial e(t)}$ from~\eqref{eq:grad_e_lp}.
Thus, the gradients to the coefficients are
\begin{equation}
    \frac{\partial \mathcal{L}}{\partial \Tilde{a}_i(t)} 
    = \frac{\partial \mathcal{L}}{\partial z_i(t)} \frac{\partial z_i(t)}{\partial \Tilde{a}_i(t)}
    = -\frac{\partial \mathcal{L}}{\partial e(t)} s(t-i).
\end{equation}

The techniques above can also be applied to time-invariant cases and are more efficient than Yu et al.~\cite{yu_singing_2023} because only one LP filter is needed while the latter requires two.
This is because we get rid of the filter for the intermediate gradients $\frac{\partial s(t)}{\partial a_i} = \text{LP}_{\bf a}(-s(t-i))$. 
We implemented time-varying LP kernels for CPU and GPU using Numba.
This reduces the runtime hundreds of times compared to a naive implementation using PyTorch operators~\cite{intro2ddsp}.

\section{Experiment}

We conducted an analysis-by-synthesis experiment to examine the end-to-end training ability of the proposed differentiable LP.
Given a voice recording, a neural encoder (analyser) predicts its time-varying latent representations (synthesis parameters) for the decoder (synthesiser), and we trained the encoder end-to-end with a simple reconstruction loss.
The decoders are made of interpretable signal-processing components (oscillators and filters).
We made our filter implementation\footnote{\url{https://github.com/DiffAPF/torchlpc}} and experiments code\footnote{\url{https://github.com/iamycy/golf}} available on GitHub.

\subsection{Dataset and training configurations}
\label{ssec:dataset}
We used the \emph{mic1} recordings from VCTK~\cite{yamagishi2019cstr} for training and evaluation.
We selected the last eight speakers as the test set, ~\emph{p225} to~\emph{p241} for validation and the rest for training.
All the recordings were downsampled to \SI{24}{\kHz}.
We follow~\cite{yu_singing_2023} to slice the training and validation data into overlapping segments with a duration of \SI{2}{\second}.
We used a batch size of 64 and trained all the encoders for 1 M steps using Adam~\cite{kingmaAdamMethodStochastic2017} with a 0.0001 learning rate.
We clipped the gradient norm to 0.5 at each step and found it effectively stabilised the training for GOLFs (Sec.~\ref{ssec:decoders}) and improved convergence for all the evaluated models.
We used the same multi-resolution spectral (MSS) loss in~\cite{yu_singing_2023} with FFT sizes $[509,1021,2053]$.
We picked the checkpoints with the lowest validation loss for evaluation.

\subsection{Speaker-independent encoder}
\label{ssec:encoder}
We extract fundamental frequency (f0) using Dio~\cite{morise_world_2016} and log-magnitude spectrograms as encoder features.
The window and hop sizes are 1024 and 240 (\SI{10}{\ms}).
Four 2D convolution layers with a kernel size $(9, 3)$ are applied to the spectrograms.
The hidden channel sizes are $[32, 64, 128, 256]$.
Each convolution is followed by batch normalisation, ReLU, and a max-pooling layer along the frequency dimension with a size of 4.
Then, the frequency and channel dimensions are flattened, concatenated with log-f0, and fed into three layers of Bi-LSTM, with a 0.1 dropout and 256 channel size.
Finally, a time-distributed linear layer, whose size depends on its corresponding decoder, converts the hidden features into synthesis parameters.
The number of parameters is 6.1 M.

\subsection{Decoders}
\label{ssec:decoders}
We trained the following four variants of GOLF:
\begin{itemize}
    \item GOLF-v1: the original HpN GOLF from~\cite{yu_singing_2023}
    \item GOLF-ss: the proposed SF GOLF in Eq.~\eqref{eq:glottal_sf}
    \item GOLF-ff: GOLF-ss but with frame-wise LP~\cite{yu_singing_2023}
    \item GOLF-fs: GOLF-ff but use sample-wise LP for inference
\end{itemize}
We linearly upsample the LPCs to the sample rate for sample-wise LP.
For all GOLFs, we oversample the signal to \SI{96}{\kHz} during wavetable synthesis (Sec.~\ref{ssec:hpn_golf}) to reduce frequency aliasing caused by linear interpolation.
We replace the cascaded biquads parameterisation with reflection coefficients~\cite{valin2019lpcnet} to tackle the responsibility problem~\cite{zhang2019fspool} in the encoder's last layer.
We use 256 frequency bins for $C(z)$.
The rest of the settings were the same as in~\cite{yu_singing_2023} except that unvoiced gating is removed for a fair comparison to the baselines.

We compared GOLFs with the following baselines: DDSP~\cite{fabbro_speech_2020}, neural homomorphic vocoder (NHV)~\cite{liu_neural_2020}, differentiable WORLD ($\nabla$WORLD)~\cite{nercessian2023differentiable}, and differentiable mel-cepstral synthesis (MLSA)~\cite{yoshimura2023embedding}, using the original configurations as closely as possible.
For NHV, we set the cepstrum order to 240 and use minimum-phase filters, as phase characteristics are not learnable through our spectral loss.
For MLSA, we set the filter order to 24 with $\alpha=0.46$.
We perform MLSA in the frequency domain with short-time Fourier transform (STFT) using \texttt{diffsptk}~\cite{sp-nitech2023sptk} as we found that training cannot converge with the Taylor-expansion-based implementation.
80 mel-frequencies are used for $\nabla$WORLD.
Band-limited pulse train is used for NHV, MLSA, and $\nabla$WORLD.
We use the DDSP implementation from~\cite{yu_singing_2023}.
All baselines use the same $C(z)$ as GOLFs, 1024 FFT size and hanning window for windowing.

\begin{figure}[h]
    \centering
    \includegraphics[width=1\columnwidth]{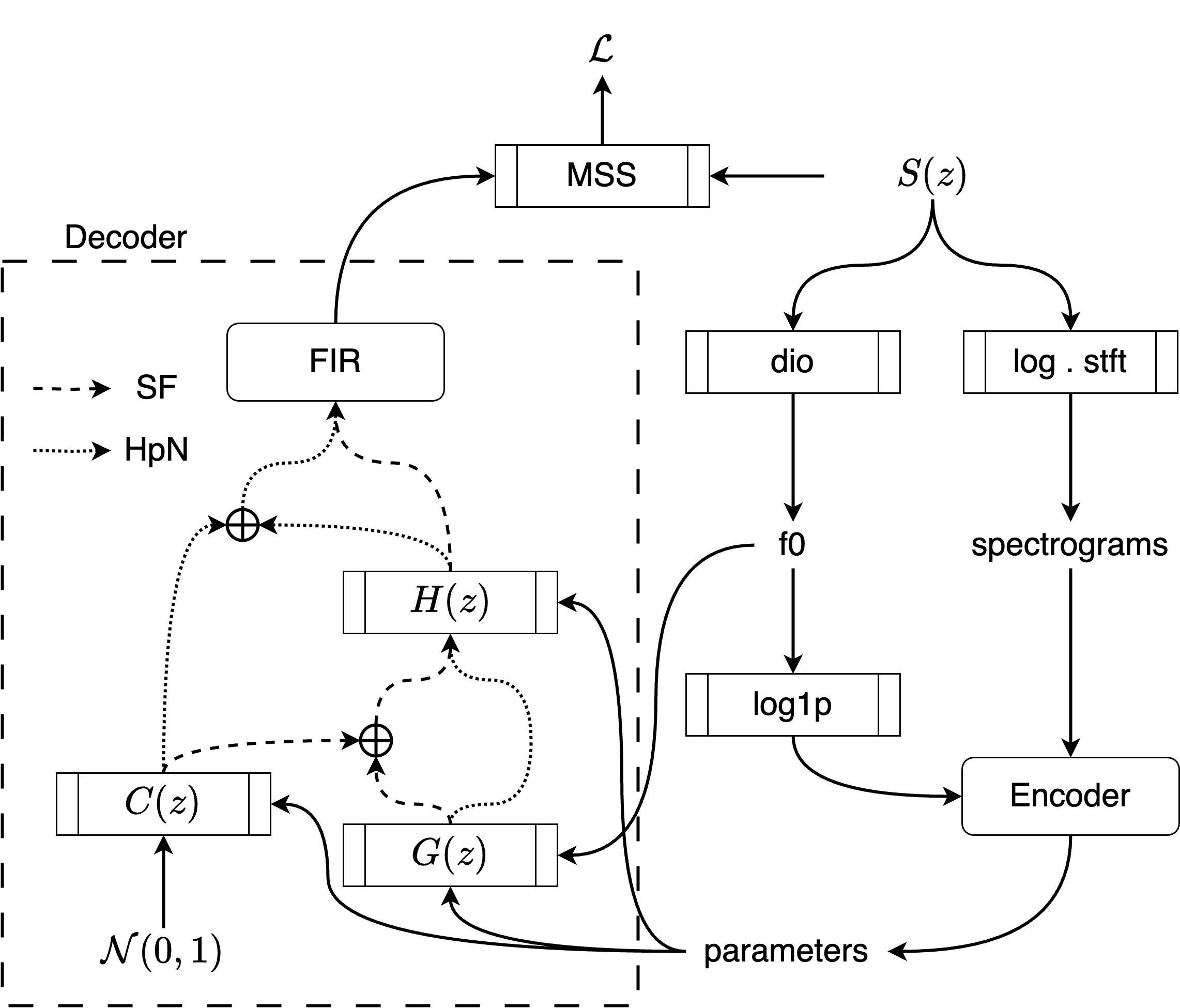}
    \caption{Flow diagram of the proposed experiment. For DDSP, the $G(z)H(z)$ is jointly modelled using an additive synthesiser.}
    \label{fig:diagram}
\end{figure}

We add a trainable FIR filter with 128 coefficients at the end of all decoders to capture the global characteristics of the VCTK dataset.
For the unvoiced speech, we sampled random oscillator frequency in Hertz from $U(50, 500)$ during training to reduce the chance of utilising harmonics for transient events~\cite{wu1_ddsp-based_2022} and used \SI{150}{\Hz} for inference.
Figure~\ref{fig:diagram} briefly illustrates the design of the whole experiment.
The training time is around 80 to 95 hours for most decoders, 121 hours for DDSP, and 135 hours for GOLF-ss, measured on an RTX A5000.

\section{Evaluations and discussions}

\begin{figure*}[t]
    \centering
    \includegraphics[width=\textwidth]{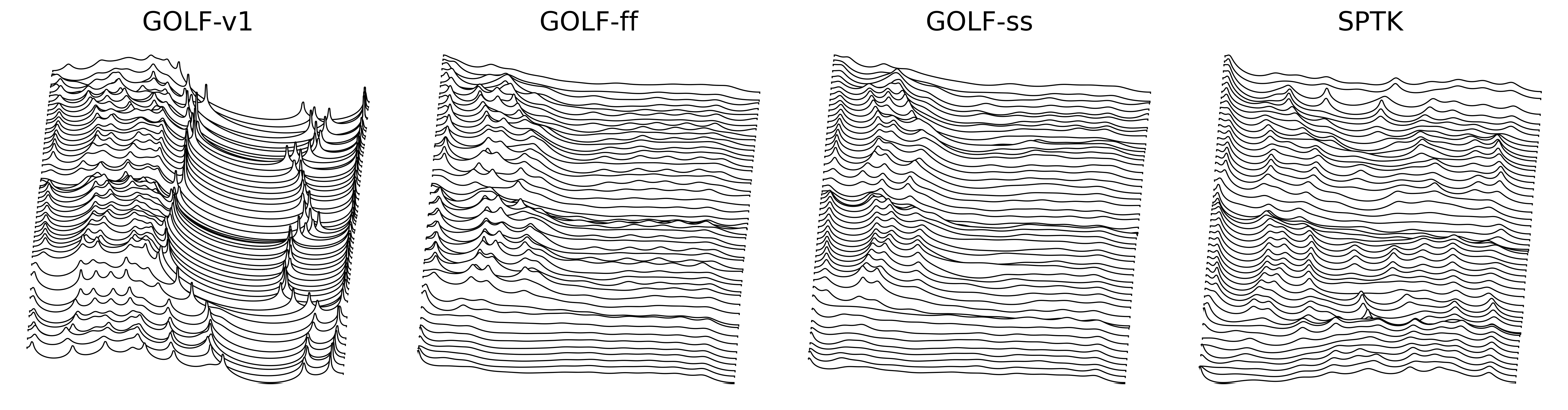}
    \caption{The running spectra converted from the encoded LPCs using 0.4 seconds of speech from speaker p360. The rightmost LPCs are computed using the auto-correlation method from SPTK with the same filter order as GOLFs.}
    \label{fig:lpc2sp}
\end{figure*}

\subsection{Objective evaluations}
\label{ssec:obj_eval}
For objective evaluations, we used MSS, mel-cepstral distortion (MCD)~\cite{kominek08_sltu}, perceptual evaluation of speech quality (PESQ), and fréchet audio distance (FAD)~\cite{kilgour19_interspeech}.
We used the same $\alpha$ as MLSA for MCD.
We computed FAD using \texttt{fadtk}~\cite{fadtk} and the embeddings from descript audio codec~\cite{kumar2023highfidelity} because it was trained on VCTK.
The FAD is calculated separately for each test speaker, and we report their mean and standard deviation; the others are averages over the test set.
We include original WORLD~\cite{morise_world_2016} as a non-differentiable baseline.

Looking at all the GOLF variants in Table~\ref{tab:speech_ae_eval}, we see adapting to SF formulation (ff) and sample-wise LP (ss) consistently improves the scores, with GOLF-ss performing the best in most metrics except FAD.
MLSA also perform competitively to GOLF-ss.
NHV has the lowest MSS loss, while $\nabla$WORLD dominates in all other metrics, indicating modelling mel-frequencies directly for vocal tract filter is likely the best for low reconstruction loss.

\subsection{Spectral analysis}
Figure~\ref{fig:lpc2sp} shows the $H(z)$ spectra of GOLFs.
GOLF-v1 shows spectral peaks with high resonance.
This could lead to unstable training and is likely why the output overflows easily when we evaluated GOLF-v1 with sample-wise LP.
We think the reason is that only the periodic signal is fed to $H(z)$, and their energies are highly centred around the harmonics.
If high resonant peaks exist between harmonics, they are not easily detectable due to a lack of energy in these regions.
There are two obvious peaks in high frequencies of GOLF-v1 spectra (with one very close to Nyquist frequency), probably due to low energy in the high frequencies of glottal pulses.
The windowing during frame-wise processing also smooths out these peaks.

Changing to SF (ff) greatly reduces this issue as the input signal includes filtered white noise, so energy is more evenly distributed.
However, the formants' position and amplitude tend to fluctuate periodically.
This problem is not obvious in the synthesised outputs due to overlap-add, as the formants in the overlapped frames are mixed. 
We use 75\% overlap in this paper.
This explains why GOLF-fs's performance deteriorates in Table~\ref{tab:speech_ae_eval}.
In contrast, GOLF-ss has the smoothest transition of formants, emphasising the importance of end-to-end training sample-wise LP.
We also see that synthesis-based LPCs are different from traditional analysis-based LPCs, due to different assumptions on the excitation signal (either flat spectra or glottal model), which is an interesting characteristic worth exploring in the future.

\begin{table}[h]
    \caption{Summary of the copy-synthesis evaluation. The values are better if they are lower, except PESQ.}
    \label{tab:speech_ae_eval}
    \centering
    \resizebox{1\columnwidth}{!}{
    \begin{tabular}{clcccc}
        \toprule
          Form. &Model &  MSS & MCD & PESQ &  FAD \\ \hline
          \multirow{3}{*}{HpN}
          &DDSP                     &  2.965&  3.42&  2.42&  32.7$\pm$7.7\\
          &NHV                      &  \textbf{2.914}&  3.32&  2.58&  31.8$\pm$7.4\\
          &GOLF-v1                  &  3.026&  3.54   & 2.36 & 39.6$\pm$9.4 \\ \hline
          \multirow{6}{*}{SF}
          &WORLD                    &  3.515&  6.07&  1.77&  270.6$\pm$56.1\\
          &MLSA                     &  3.006&  3.35&  2.48&  40.1$\pm$10.0\\
          &$\nabla$WORLD            &  2.918&  \textbf{3.26}&  \textbf{2.66}&  \textbf{22.4}$\pm$\textbf{5.6}\\
          &GOLF-ss                  &  3.005&  3.43&  2.49&  38.4$\pm$9.2\\
          &GOLF-ff                  &  3.011&  3.46&  2.39&  34.0$\pm$7.7\\
          &GOLF-fs                  &  3.074&  3.70&  2.16&  44.1$\pm$10.1\\
        \bottomrule
    \end{tabular}
    }
\end{table}

\subsection{Subjective evaluation}

Based on the results from Table~\ref{tab:speech_ae_eval}, we picked GOLF-ss, NHV, and $\nabla$WORLD (represent the best models for GOLFs/HpN/SF, respectively) for a MUSHRA listening test.
We selected \emph{p360} (male) and \emph{p361} (female) from the test set as they have the lowest FAD score on average.
We picked ten different utterances and randomly assigned five to each speaker.
The duration of the audio ranged from 5 to 7 seconds.
Each test example consists of re-synthesised audio by the selected models and a low anchor model, using the same utterance.
The low anchor is pulse trains with traditional LPC analysis.
The ground truth recording is included as a hidden reference.

The test was conducted on Go Listen~\cite{hines_go_2021}.
We sent out the test to the mailing lists of related research communities.
A different utterance from \emph{p360} was used to train the participant before the test, using the low anchor and the ground truth.
We asked the participants to rate the audio quality on a scale from 0 to 100 and encouraged them to use the full scale for each example as much as possible.
The orders of the examples and models were randomised.
Among 21 participants, we excluded four as they did not consistently rate the low anchor the lowest.
According to self-reports, all the remaining participants used headphones.

\begin{figure}
    \centering
    \begin{tikzpicture}
    \begin{axis}[
        x tick label style={font=\footnotesize},
        enlarge x limits=0.12,
        ylabel=MUSHRA,
        xtick=data,
        symbolic x coords = {Anchor, NHV, $\nabla$WORLD, GOLF-ss, Reference},
        legend style={at={(0.22,0.94)}, anchor=north,legend columns=-1},
        ybar,
        ymin=0,
        ymajorgrids=true,
    ]
    \addplot+[
        error bars/.cd,
        y dir=both,
        y explicit,
    ] coordinates {
        (Anchor, 6.51764706) +- (0, 0.95943094)
        (NHV, 37.62352941) +- (0, 3.10652532)
        ($\nabla$WORLD,44.5294117) +- (0, 3.858009)
        (GOLF-ss,58.52941176) +- (0, 3.87094237)
        (Reference,88.68235294) +- (0, 2.65150639)
    };
    \addplot+[
        error bars/.cd,
        y dir=both,
        y explicit,
    ] coordinates {
        (Anchor, 8.63529412) +- (0, 1.30800536)
        (NHV, 37.23529412) +- (0, 3.18823523)
        ($\nabla$WORLD,46.83529412) +- (0, 4.12873333)
        (GOLF-ss,46.89411765) +- (0, 3.74278497)
        (Reference,86.84705882) +- (0, 2.49707063)
    };
     
    \legend{p360,p361}
    \end{axis}
    \end{tikzpicture}
    \caption{The average ratings of each speaker with a 95\% confidence interval.}
    \label{fig:mos}
\end{figure}
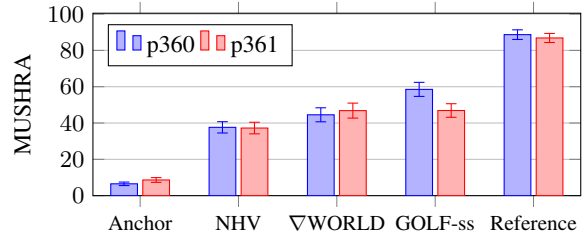

The result is shown in Figure~\ref{fig:mos}.
For speaker \emph{p360}, GOLF significantly outperforms the others.
After inspecting the samples, we found NHV and $\nabla$WORLD have a slight but audible robotic timbre.
We hypothesise that the use of pulse trains largely contributes to this issue.
In contrast, we do not see the same trend in ratings for speaker \emph{p361}.
We found that the pitch estimator (Dio) performs poorly on \emph{p361}, with more misclassification of voiced/unvoiced signals than \emph{p360}.
This leads to breathy voice artefacts, whose effect surpasses the robotic timbre issue, decreasing overall ratings and making the comparison of GOLF and $\nabla$WORLD harder.
One possible solution is switching to a more accurate pitch estimator during training or evaluation.

\section{Conclusions and future work}
In this paper, we derived and implemented efficient gradient backpropagation for differentiable time-varying LP, which helps train sample-rate level LP filters end-to-end in a reasonable time.
In the copy-synthesis evaluation, we show that changing the vocoder to source-filter form or replacing frame-wise LP approximation with the proposed implementation both help the model learn smoother LPCs.
Nevertheless, more work is needed to analyse the learnt synthesis parameters and the cause of robotic timbre when using pulse trains as periodic sources, which constitute future work.

\section{Acknowledgements}
We thank Takenori Yoshimura for giving feedback on our early drafts.
The first author is a research student at the UKRI Centre for Doctoral Training in AI and Music, supported jointly by UKRI [grant number EP/S022694/1] and Queen Mary University of London.

\bibliographystyle{IEEEtran}
\bibliography{clean}

\end{document}